\documentclass[sigconf, screen]{acmart}
\AtBeginDocument{%
  \providecommand\BibTeX{{%
    \normalfont B\kern-0.5em{\scshape i\kern-0.25em b}\kern-0.8em\TeX}}}

\usepackage{enumitem}
\usepackage[list=true]{subcaption}
\usepackage{tabularx}
\usepackage{float}
\usepackage{stfloats}
\usepackage[ruled, lined, linesnumbered, commentsnumbered]{algorithm2e}
\let\oldnl\nl
\newcommand{\nonl}{\renewcommand{\nl}{\let\nl\oldnl}}

\usepackage{xcolor}
\usepackage[T1]{fontenc}
\usepackage{amsmath}
\usepackage{cleveref}
\usepackage{amsthm}
\usepackage{amsfonts}

\usepackage{amssymb}
\usepackage{bbm}
\usepackage{dsfont}
\usepackage{multirow}
\usepackage{diagbox}
\usepackage{graphicx}
\usepackage{ifthen}
\graphicspath{{/.images/}}

\newcounter{casenum}

\DeclareMathOperator*{\argmax}{arg\,max}

\newcommand{\ignore}[1]{}

\newboolean{showcomments} \setboolean{showcomments}{true}

\newcommand{\vd}[1]{\ifthenelse{\boolean{showcomments}}
  {\textcolor{violet}{(VD says: #1)}} {} }
\newcommand{\mo}[1]{\ifthenelse{\boolean{showcomments}}
  {\textcolor{blue}{(MH says: #1)}} {} }
\newcommand{\jt}[1]{\ifthenelse{\boolean{showcomments}}
  {\textcolor{teal}{(Jia says: #1)}} {} }
\newcommand{\trs}[1]{\ifthenelse{\boolean{showcomments}}
  {\textcolor{teal}{(Tanay says: #1)}} {} }

\begin{document}

\title{Degradation-Aware Frequency Regulation of a Heterogeneous Battery Fleet via Reinforcement Learning}

\author{Tanay Raghunandan Srinivasa}
\affiliation{%
  \institution{Plaksha University}
  \city{Mohali}
  \country{India}}
\email{tanay.srinivasa@plaksha.edu.in}
\author{Vivek Deulkar}
\affiliation{%
  \institution{Plaksha University}
  \city{Mohali}
  \country{India}}
\email{vivek.deulkar@plaksha.edu.in}
\author{Jia Bhargava}
\affiliation{%
  \institution{Plaksha University}
  \city{Mohali}
  \country{India}}
\email{jia.bhargava@plaksha.edu.in}
\author{Mohammad Hajiesmaili}
\affiliation{%
  \institution{University of Massachusetts Amherst}
  \city{Amherst}
  \country{United States of America}}
\email{hajiesmaili@cs.umass.edu}
\author{Prashant Shenoy}
\affiliation{%
  \institution{University of Massachusetts Amherst}
  \city{Amherst}
  \country{United States of America}}
\email{shenoy@cs.umass.edu}

\renewcommand{\shortauthors}{Srinivasa, et. al}
\acmSubmissionID{409}

\begin{abstract}
Battery energy storage systems are increasingly deployed as fast-responding resources for grid balancing services such as frequency regulation and for mitigating renewable generation uncertainty. However, repeated charging and discharging induces cycling degradation and reduces battery lifetime. This paper studies the real-time scheduling of a heterogeneous battery fleet that collectively tracks a stochastic balancing signal subject to per-battery ramp-rate and capacity constraints, while minimizing long-term cycling degradation.

Cycling degradation is fundamentally path-dependent: it is determined by charge--discharge cycles formed by the state-of-charge (SoC) trajectory and is commonly quantified via rainflow cycle counting. This non-Markovian structure makes it difficult to express degradation as an additive per-time-step cost, complicating classical dynamic programming approaches. We address this challenge by formulating the fleet scheduling problem as a Markov decision process (MDP) with constrained action space and designing a dense proxy reward that provides informative feedback at each time step while remaining aligned with long-term cycle-depth reduction.

To scale learning to large state--action spaces induced by fine-grained SoC discretization and asymmetric per-battery constraints, we develop a function-approximation reinforcement learning method using an Extreme Learning Machine (ELM) as a random nonlinear feature map combined with linear temporal-difference learning. We evaluate the proposed approach on a toy Markovian signal model and on a Markovian model trained from real-world regulation signal traces obtained from the University of Delaware, and demonstrate consistent reductions in cycle-depth occurrence and degradation metrics compared to baseline scheduling policies.

\end{abstract}

\acmYear{}\copyrightyear{}
\setcopyright{acmcopyright}
\acmConference[]{}{}{}
\acmBooktitle{}
\acmPrice{}
\acmDOI{}
\acmISBN{}



\maketitle

\section{Introduction}\label{sec:intro}
Power system operators increasingly rely on fast-responding energy storage resources to provide frequency regulation and other ancillary services. Battery energy storage systems (BESSs), in particular, are well suited for frequency regulation due to their rapid ramping capabilities, high round-trip efficiency, and modular deployability \cite{luo2015overview}. In practice, frequency regulation is often provided not by a single large battery, but by an aggregated fleet of heterogeneous battery units that collectively track a regulation signal provided by the system operator \cite{liu2013decentralized,janjic2017commercial}.

Despite their operational advantages, batteries suffer from cycling-induced degradation \cite{bat_degradation1,bat_deg_Millner}, which significantly affects their lifetime and long-term economic viability. A central challenge in operating battery fleets for frequency regulation is therefore to balance regulation performance with degradation minimization. This challenge is compounded by two fundamental characteristics of the problem. First, battery fleets are heterogeneous: individual units may differ in energy capacity, ramping limits, and current state of charge (SoC). Second, degradation is path-dependent—it depends not only on instantaneous charge or discharge actions, but on the entire sequence of actions that shape the SoC trajectory over time.

In this work, we study the problem of operating a fleet of $N$ heterogeneous battery units to provide frequency regulation while minimizing long-term cycling degradation. The regulation signal is modeled as a discrete-time stochastic process evolving over a finite state space, capturing the uncertainty and temporal correlation inherent in regulation demand. At each time slot, the aggregator must decide how to distribute the regulation request across the battery units, subject to per-battery energy capacity and ramping constraints, as well as collective feasibility constraints. These decisions determine the evolution of each battery’s SoC and, consequently, the cycle formation patterns that drive degradation.

A key difficulty in this setting is that battery degradation cannot be expressed as a simple additive cost per time step. Instead, degradation is commonly modeled using cycle-counting methods—such as the rainflow counting algorithm \cite{downing1982simple}—which identify charge-discharge cycles from SoC trajectories and assign damage based on cycle depth. Even though such models are accurate (see \cite{bat_deg_Millner,bat_degradation1}), they are inherently non-Markovian and do not naturally admit an incremental per-step cost representation. This poses a major obstacle to applying standard optimization and control techniques, including dynamic programming.

To address this challenge, we adopt a reinforcement learning (RL) framework. RL is particularly well suited to this problem for multiple reasons. The regulation signal dynamics are modeled as a Markov process with unknown transition probabilities, making model-free learning attractive. The control decisions influence degradation through long-term SoC trajectories rather than instantaneous effects, requiring a method capable of reasoning over extended horizons. Also, the action space is state-dependent and constrained by physical battery limits, further complicating classical approaches.

However, applying RL in this context introduces its own challenges. One of the major challenges is to design a reward function that incentivizes the agent to shape SoC trajectories in a degradation-aware manner, despite the delayed and path-dependent nature of true cycling damage. Directly using cycle-based degradation as a reward leads to sparse and delayed reward, which is poorly suited for learning. Consequently, an appropriate reward-shaping strategy is required—one that provides informative intermediate feedback while remaining aligned with long-term degradation minimization.

In this paper, we propose a degradation-aware RL framework for frequency regulation using heterogeneous battery fleets. We formulate the problem as a Markov decision process (MDP) and introduce a dense proxy reward design derived from real-time properties of SoC trajectories that encourages shallow cycling and avoids deep cycle discharge. The learned policy is evaluated using the full rainflow cycle counting to quantify the actual degradation, ensuring that performance is assessed using a physically meaningful metric.

To scale learning to large state and action spaces, we employ a function approximation to estimate $Q$-functions. Specifically, we use an Extreme Learning Machine (ELM) as a random nonlinear feature map over state--action pairs, and learn a linear value function using semi-gradient temporal-difference updates. This approach enables efficient learning while retaining nonlinear representational capabilities.
\subsection*{Our Contributions}
The main contributions of this paper are as follows:
\begin{itemize}
    \item We formulate the operation of heterogeneous battery fleets for frequency regulation as a discrete-time MDP with stochastic regulation demand, explicit ramping and capacity constraints, and long-term degradation objectives.
    \item We develop a dense reward-shaping mechanism that guides the RL agent toward SoC trajectories with reduced cycling depth, while remaining aligned with cycle-based degradation models that detect cycles and associated degradation only over long time horizons.
    \item We propose an RL solution based on random feature extraction using Extreme Learning Machines combined with linear temporal-difference learning
for value-function approximation.
    \item We evaluate the proposed approach using realistic regulation signals and quantify performance using rainflow-based degradation metrics, demonstrating significant reductions in cycling damage compared to baseline dispatch strategies.
\end{itemize}

\subsection*{Related Work}

Data-driven methods, and reinforcement learning (RL) in particular, have
made substantial progress in solving sequential decision-making problems.
The introduction of neural networks as function approximators in RL,
popularized by the seminal work in~\cite{mnih2015human},
demonstrated that model-free learning can scale to large and complex
state spaces. Since then, RL has been widely applied in energy systems,
including demand response \cite{vazquez2019reinforcement}, energy management, and storage control such as electric vehicle charging scheduling~\cite{wan2018model}
and energy arbitrage using battery storage~\cite{wang2018energy,wankmuller2017impact}.
While
these studies demonstrate the flexibility of RL in handling uncertainty
and complex system dynamics, battery degradation is either ignored altogether or typically modeled
in a simplified manner and do not capture
the cycle-dependent aging behavior that is critical for applications
involving frequent charge--discharge operation, such as frequency
regulation.


The works in \cite{bat_deg_Millner,bat_degradation1} rigorously study battery degradation using both physics-based and data-driven techniques, offering models that account for cycle-based degradation. \cite{zhang2020identifying} uses machine learning to identify degradation patterns.

There are recent work that has tried to bridge the gap between RL-based control and
cycle-based degradation \cite{KWONZHU9789478,bat_deg_via_RL,hao2023adaptive,zhang2024load}. \cite{KWONZHU9789478}
formulate frequency regulation with battery storage as an MDP in which
regulation tracking performance and cycling degradation are combined into
an additive cost function. Their approach augments the state with multiple
recent switching points of the state-of-charge (SoC) trajectory to infer
cycle formation. In contrast, our formulation enforces frequency regulation
requirements as hard feasibility constraints and relies only on the most
recent switching point to construct a degradation-aware reward, enabling
simpler state representation and online implementation. The work in \cite{bat_deg_via_RL} incorporate battery degradation into RL by
introducing a linearized approximation of cycle-based degradation as an
instantaneous per-step cost. However, this requires the entire episode data trace for linearization and must be updated episodically. This sensitivity to the
chosen trajectory can lead to modeling mismatch and may limit the
ability of the learning algorithm to explore and converge to optimal
policies over the full policy space.

Overall, while prior work has demonstrated the potential of reinforcement
learning for energy storage operation and has explored various approaches
to modeling battery degradation, effectively incorporating accurate,
cycle-based degradation into scalable, online RL frameworks for frequency
regulation remains an open problem. The present work addresses this gap
by designing a degradation-aware reward that is compatible with real-time
learning and by developing a stable function-approximation-based RL
approach for a heterogeneous battery fleet.
\section{Model and Problem Formulation}\label{sec:model}
We consider the operation of an aggregator controlling a fleet of $N$ heterogeneous battery energy storage units to provide frequency regulation over a discrete-time horizon. Time is slotted as 
$t=0,1,2,\dots$ with each slot corresponding to a fixed duration $\Delta.$
\subsection{Frequency Regulation Signal Model}
Let $r(t) \in \mathbbm{Z}$ denote the frequency regulation request at time slot $t,$ measured in units of energy over the slot duration. Positive values of $r(t)$ correspond to a request for net charging (absorbing energy from the grid), while negative values correspond to a request for net discharging (injecting energy into the grid). Due to fine-grained energy measurement and discretization, all quantities in the system—including battery energy levels, ramping limits, and regulation requests—are assumed to be integer-valued.

The regulation signal is modeled as an exogenous stochastic process with unknown dynamics. To obtain a finite-state MDP amenable to reinforcement learning, we approximate the signal evolution using a discrete-time Markov chain (DTMC) over a finite state space $\mathcal{S}_r,$ with unknown transition probabilities. This model captures both the stochastic nature of regulation demand and its temporal correlation, while allowing for model-free learning.

\subsection{Battery Model and Local Constraints} \label{subsec:battery-model}

Each battery unit $i\in\{1,\ldots,N\}$ is characterized by the following parameters:
energy capacity $B_i\in\mathbb{Z}_+$, maximum charging rate $c_i\in\mathbb{Z}_+$, maximum discharging rate $d_i\in\mathbb{Z}_+$. Recall that we measure energy with fine granularity and, consequently, all of these quantities are represented as integer values. This discretization does not affect modeling fidelity.


Let $b_i(t)\in\{0,1,\ldots,B_i\}$ denote the state of charge (SoC) of battery $i$ at the beginning of time slot $t$.
During slot $t$, the aggregator agent selects an action$a_i(t)\in\mathbb{Z}$, where $a_i(t)>0$ represents
charging and $a_i(t)<0$ represents discharging. The per-battery action is constrained by both ramping limits and energy availability:
\begin{equation}\label{eq:ramp and feas constraints}
-\min\{d_i,\,b_i(t)\}\ \le\ a_i(t)\ \le\ \min\{c_i,\,B_i-b_i(t)\},
\end{equation}
where $c_i,d_i>0$ represent maximum possible charging and discharging in one time step.
The SoC then evolves deterministically according to
\begin{equation}\label{eq:battery_dynamics}
b_i(t+1)=b_i(t)+a_i(t),
\end{equation}
ensuring that the battery energy level never exceeds its capacity or falls below zero.

\subsection{Collective Feasibility and Regulation Tracking}

At each time slot, the battery fleet must collectively track the regulation request as closely as possible, subject to physical
feasibility. We define the aggregate feasible charging and discharging limits at time $t$ as
\begin{equation} \label{eq:action_constraint1}
A_{\max}(t)=\sum_{i=1}^{N}\min\{c_i,\,B_i-b_i(t)\},\qquad
A_{\min}(t)=-\sum_{i=1}^{N}\min\{d_i,\,b_i(t)\}.
\end{equation}

The maximum regulation that can be served by the fleet is therefore bounded within
$[A_{\min}(t),A_{\max}(t)]$. We define the served regulation
\begin{equation}
\tilde r(t)=\mathrm{clip}\bigl(r(t),A_{\min}(t),A_{\max}(t)\bigr),
\end{equation}
where
\begin{equation}\label{eq:clip}
\mathrm{clip}(x,\ell,u)=\min\{\max\{x,\ell\},u\}.
\end{equation}

The action vector $a(t)=(a_1(t),\ldots,a_N(t))$ must satisfy the collective constraint
\begin{equation}\label{eq:regulation_constraint}
\sum_{i=1}^{N} a_i(t)=\tilde r(t),
\end{equation}
in addition to the per-battery constraints described in Section \ref{subsec:battery-model}. The set of admissible actions at state $s(t)$ is therefore
state-dependent and denoted by $\mathcal{A}_{s(t)}$.

This formulation ensures that the fleet provides the maximum feasible regulation at every time step, while allowing the learning
agent to decide how the regulation burden is distributed across individual batteries.

We consider a scenario involving multiple battery storage units\footnote{We use the terms battery and storage unit interchangeably in this paper} operated in the presence of a stochastic frequency regulation signal $r(t).$ When the regulation signal is positive, the storage units must collectively absorb $|r(t)|$ units of energy; when the signal is negative, they must collectively discharge $|r(t)|$ units of energy to provide the required regulation. The repeated charging and discharging of the storage units leads to degradation of battery health, which in turn reduces their operational lifetime. This degradation is commonly characterized in terms of cycling behavior, with battery lifetime measured by the number of charge–discharge cycles that can be sustained before the usable capacity degrades beyond acceptable limits.

The goal is to intelligently schedule these multiple storage units, i.e., to carry out their charge-discharge mechanism, so that the the regulation $r(t)$ is distributed across these $N$ storage units and it results in the least possible degradation.

\subsection{Cycling Degradation Model}
\label{sec:bat_deg_model}
Battery degradation due to charge–discharge cycling is highly dependent on the depth and frequency of SoC cycle that gets formed. A widely used and experimentally validated approach to quantifying cycling degradation is the rainflow cycle counting algorithm, which decomposes an SoC trajectory into charge–discharge cycles and assigns damage based on each cycle’s depth of discharge (DoD)\footnote{Although battery degradation is influenced by several factors, including temperature, C-rate, average state of charge, battery chemistry, and electrochemical aging mechanisms, prior studies have shown that cycling behavior is a primary driver of degradation in applications involving frequent charge–discharge operation \cite{bat_degradation1}. Therefore, we restrict attention to cycle-induced degradation in this work, while acknowledging that accounting for the comprehensive degradation modeling lies beyond the scope of this paper.}.

Let $\Phi(b_i(\cdot))$ denote the set of cycles identified from the SoC trajectory $b_i(\cdot)$ of battery $i,$ and let $\delta$ denote the DoD of a given cycle. The degradation incurred by that cycle is modeled using a stress function $f(\delta),$ which is typically nonlinear and increasing in $\delta.$ We adopt the battery degradation model in \cite{bat_deg_Millner,shi2018convex} where the stress function $f(\delta)$ takes a polynomial form as 
$f(\delta)= \alpha e^{\beta \delta} \in (0,1),$
where $\alpha$ and $\beta$ are constants which are specific to the battery chemistry and can be inferred from the battery data sheets provided by the battery manufacturer. The total cycling degradation of battery $i$ over a horizon $T$ is given by

\begin{equation}\label{eq:tot_degradation_in_bat_i}
    D_i(T)=\sum_{\delta \in \Phi(b_i(\cdot))} f(\delta).
\end{equation}

The fleet-wide degradation is then $
    D(T)=\sum_{i=1}^T D_i(T).
$ While tractable, this degradation model is inherently non-Markovian, as cycle identification depends on the entire SoC trajectory rather than on instantaneous state transitions. This presents a fundamental challenge for control and learning: cycles may overlap and complete at different time instants, making degradation a delayed and path-dependent quantity. 

\subsection{Control Objective}
The system objective is to operate the batteries over a time horizon $T$ so as to minimize the total cycling degradation while providing frequency regulation:
\begin{equation}\label{eq:cost_fun}
    \min_{\{a(t)\}_{t=0}^{T-1}} \sum_{i=1}^N L_i
\end{equation}
subject to SOC dynamics \eqref{eq:battery_dynamics}, individual battery constraints \eqref{eq:action_constraint1} and \eqref{eq:regulation_constraint}, collective regulation constraint \eqref{eq:regulation_constraint}.
The decision variables are the sequence of action vectors $a(0), a(1), \dots, a(T-1), $ wherein action vector $a(t)=(a_1(t), \cdots, a_N(t))$ satisfies action constraints \eqref{eq:ramp and feas constraints} and \eqref{eq:regulation_constraint} during every time step. 

\subsection{Why a Sequential Decision-Making Framework Is Needed}

Although the objective in \eqref{eq:cost_fun} is well defined, it is not amenable to greedy optimization. The degradation incurred by a battery depends on entire SoC trajectories, not on instantaneous actions. An action that appears benign in the short term may contribute to deep cycles later, while a slightly more aggressive action now may prevent large cycles in the future.

This delayed and history-dependent structure makes the problem inherently sequential and motivates a Markov Decision Process (MDP) formulation, which is developed in the next section.\footnote{Note that even though the cycle formation is history dependent and hence non-Markovian, we analysed the problem through the MDP framework by appropriately designing a surrogate reward function that can capture the non-Markovian, history-dependent nature of cycle formation (see details in Section~3).}


Since the degradation is cycle based, it is not immediately clear how the current choice of action $a(t)$ affects the SoC cycle formation across the $N$ batteries in the future. Consequently, any greedy policy, which choose the action with least associated degradation, will only consider whether or not an immediate cycle(s) is getting formed in the current time slot by the chosen action when it applied. Such policies are myopic and completely ignores the foresightedness of current good (greedy) action being detrimental in the future if it leads to large cycling degradation.   

This motivates us go for Markov decision processes (MDP) approach where such temporal dependence of action on the reward obtained is well addressed (see \cite{sutton2018reinforcement}).
The MDP model is discussed in the next subsection.

\section{Degradation-Aware Learning Framework}\label{sec:rl_framework}


In this section, we cast the coordinated battery scheduling problem as a Markov Decision Process (MDP) and design a reward function that incentivizes long-term degradation-aware behavior. The MDP formulation enables the learning agent to account for the temporal coupling between present actions and future battery cycling behavior.
\subsection{MDP model} \label{subsec:mdp-model}

We consider an infinite-horizon, discounted MDP defined by the tuple
$(\mathcal{S},\mathcal{A},\mathcal{P},\mathcal{R}, \gamma),$
where $\mathcal{S}$ is the state space, $\mathcal{A}$ is the action space, $\mathcal{P}$ denotes the state transition kernel, $\mathcal{R}$ is the reward function, and $\gamma \in (0,1)$ is the discount factor.

\paragraph{State Space} The system state at time $t$ is defined as
 $$s(t)=(b_1(t),b_2(t),\dots,b_N(t),r(t)) \in \mathcal{S},$$ which captures the SoC of all batteries and the current regulation request. Each battery state 
$b_i(t)$ takes integer values in 
$\{0,1,\dots, B_i\}$ and the regulation signal $r(t)$ takes values in the finite set $\mathcal{S}_r.$ Since energy is measured with fine granularity, discretization introduces no loss of modeling fidelity while ensuring a finite state space. The full state space is given by
$
\mathcal{S} = \prod_{i=1}^{N} \{0,1,\ldots,B_i\} \times \mathcal{S}_r,
$
with cardinality
$
|\mathcal{S}| = \left( \prod_{i=1}^{N} (B_i + 1) \right) |\mathcal{S}_r|.
$

\paragraph{Action Space}
 An action at time $t$ is the vector
 $a(t)=(a_1(t), \cdots, a_N(t))$ where $a_i(t) \in \mathbbm{Z}$ represents the energy charged to or discharged from battery $i$ during time slot $t.$ The set of feasible actions $A_{s(t)} \in \mathbbm{Z}^N$ is state dependent and consists of all action vectors satisfying: individual ramp-rate and capacity constraints (1)–(3), the collective regulation tracking constraint (6). The state dependence of $\mathcal{A}_{(\cdot)}$ significantly increases the complexity of the problem, as the number of feasible actions varies with the SoC configuration.

 \paragraph{State Transitions}
 The state transition mechanism has two components\begin{enumerate}
     \item Battery dynamics: Given $s(t)$ and $a(t),$ the battery SoC evolves deterministically via (2)
     \item Regulation signal dynamics: The regulation signal $r(t)$ evolves as a DTMC over state space $\mathcal{S}_r$ with unknown transition probabilities. 
 \end{enumerate}
  Thus, the transition kernel 
$P\left(s(t+1)|s(t),a(t)\right)$ is Markovian but unknown, making model-free RL appropriate.

 \paragraph{Reward function} The MDP reward $R\left(s,a,s^\prime\right)$ is the reward that the environment offers to the agent when the environment transitions from $s$ to $s^\prime$ under the influence of action $a$ taken.  
The reward structure has the following characteristics: The immediate reward is designed to reflect the long-term objective of minimizing cycling degradation. However, since true degradation cannot be computed incrementally, the reward function is constructed using a proxy based on real-time properties of SoC trajectories. 
The detailed reward design is presented in Section~\ref{sec:MDP_reward_design}. 

\subsection{Objective and Value Function}
In the MDP setting, the objective of the aggregator is to find a policy $\pi$ that selects actions $a(t)=\pi(s(t))$ at each time $t$ so as to minimize the expected long-term degradation, by maximizing the MDP discounted return:
\begin{equation*}
    \max_{\pi} \mathbbm{E}\Big[ \sum_{t=0}^\infty  \gamma^t R\big(s(t), a(t), s(t+1) \big) \Big].
\end{equation*}
The corresponding state–action value function (Q-function) under policy 
$\pi$ is defined as
\[
Q^{\pi}(s,a)
= \mathbb{E}_{\pi}\!\left[
\sum_{t=0}^{\infty} \gamma^{t}
R\bigl(s(t), a(t), s(t+1)\bigr)
\;\middle|\;
s(0)=s,\; a(0)=a
\right],
\]
 which expresses long-term performance as the immediate reward plus the discounted optimal value of the next state. 

As discussed in Section \ref{sec:model}, the true objective of the aggregator is to minimize long-term cycling degradation, which depends on charge–discharge cycles identified from the SoC trajectories of individual batteries. However, as mentioned, cycle-based degradation models such as rainflow counting are inherently non-Markovian and do not admit a natural per-time-step cost representation. This poses a challenge in designing an appropriate MDP reward function that assigns a reward (or penalty) at each time step.

\subsection{Challenges in MDP Reward Design}
Because cycle-based degradation is defined only upon cycle completion, an ideal reward function $R(s,a,s')$ would assign a degradation penalty to action $a$ only for those transitions $(s,a,s')$ where a cycle-counting procedure (such as rainflow counting algorithm) identifies a completed cycle and computes its depth of discharge; otherwise $R(s,a,s') = 0$. Formally, if a degradation cost $f(\delta)$ is incurred only when a cycle of depth $\delta$ is identified, then the instantaneous reward is zero for most time steps and nonzero only at cycle-completion events.

Such a reward structure yields a sparse and delayed learning signal, since penalties arrive only after extended sequences of actions that collectively form a cycle. Learning under this type of feedback is computationally inefficient and unstable in reinforcement learning, particularly in large state spaces. These observations motivate the design of a dense proxy reward that provides informative feedback at every time step while remaining aligned with the long-term objective of reducing cycling degradation.

\subsection{Degradation-Aware MDP Reward Design} \label{sec:MDP_reward_design}

To design a dense reward that reflects long-term cycling degradation,
we exploit structural properties of charge--discharge cycles formed
by state-of-charge (SoC) trajectories. Cycle-based degradation models,
such as rainflow counting, identify cycles between successive local
maxima and minima of the SoC profile. However, cycles of different
depths may overlap in time: a large-amplitude cycle can complete and
be removed while a smaller-amplitude cycle, initiated earlier,
remains unclosed. Figure~1 illustrates such overlapping and nested
cycle formations.

This observation motivates tracking the progression of \emph{ongoing}
cycles using information that can be computed online, without access
to future SoC values. In particular, we use \emph{switching points}\footnote{A switching point (or turning point) is the SOC value that defines either a local peak or a local valley in the SOC profile.}, which correspond to local extrema of the SoC
trajectory and serve as reference points for measuring the depth of
incomplete cycles.

For each battery $i$, let $b_i^{\mathrm{SP}}(t)$ denote the most recent
switching point of its SoC trajectory prior to time $t$. If the SoC is
currently increasing, $b_i^{\mathrm{SP}}(t)$ corresponds to the most
recent local minimum; if the SoC is decreasing, it corresponds to the
most recent local maximum. Switching points can be detected and
updated incrementally by monitoring changes in the direction of SoC
evolution, making them suitable for real-time learning and control.

For each battery $i,$ the switching point $b^{SP}_i(t)$ is defined as the most recent local extremum of its SoC trajectory prior to time $t.$ Specifically, if the SoC trajectory is currently increasing, the switching point corresponds to the most recent local minimum; if the SoC trajectory is currently decreasing, the switching point corresponds to the most recent local maximum. Switching points can be updated incrementally by monitoring changes in the direction of SoC movement, without requiring access to future information. This makes them well suited for real-time learning and control. 

The quantity $|b_i(t)-b^{SP}_i(t)|$ represents the current magnitude of SoC deviation from the most recent switching point for battery $i.$ This deviation serves as a key indicator of the depth of the charge–discharge cycle that is currently being formed.

We now construct a dense reward function that penalizes large deviations of the SoC from switching points and thereby encourages shallow cycling behavior. For each battery $i,$ we define the instantaneous degradation proxy reward as

\begin{equation}\label{eq:reward_fun}
      r_i(t) = - \Big( \sum_{i=1}^N \alpha_d e^{\beta|b_i(t)+a_i(t)-b^{SP}_i(t)|} - \alpha_d e^{\beta|b_i(t)-b^{SP}_i(t)|} \Big ),
\end{equation}
where $\alpha, \beta$ are the degradation parameters.
The MDP one step reward at time $t$ is given by
\begin{equation}\label{eq:MDP_reward}
R\big(s(t), a(t), s(t+1) \big) =\sum_i^{N} r_i(t).
\end{equation}
The negative sign reflects the interpretation of the reward as a
degradation penalty: maximizing cumulative reward corresponds
to minimizing cumulative degradation. 
This reward-shaping mechanism incentivizes the learning agent to distribute regulation effort across batteries in a manner that avoids deep SoC deviations in any single unit, thereby reducing the formation of high-depth cycles over time.

Note that the proxy reward does not compute true cycling degradation. Instead, it serves as a surrogate objective whose minimization empirically leads to SoC trajectories with lower rainflow-counted degradation. In other words, the objective of the MDP is defined in terms of a proxy reward that is correlated with, but not identical to, cycle-based degradation. True cycle-based degradation is computed only during evaluation of the trained RL agent, as described and reported in simulation Section~\ref{sec:exp_evaluation}.

\subsection*{Connection of MDP Reward to Cycle-Based Degradation}

A switching point (or turning point) in an SoC trajectory corresponds to a local
maximum or minimum and marks a change in the direction of SoC evolution. In
cycle-counting methods such as rainflow counting, completed charge–discharge
cycles are identified between successive switching points, and the depth of
discharge (DoD) of a cycle is determined by the difference between the associated
extrema. Intuitively, the quantity $|b_i(t)-b^{SP}_i(t)|$ tracks the partial depth of a developing cycle. Penalizing growth in this quantity intuitively encourages the controller to reverse direction earlier, distribute regulation across batteries, avoid concentrating large excursions on a subset of units. As a result, the learned policy implicitly shapes the SoC trajectories to reduce the frequency and depth of cycles detected by rainflow counting.




\subsection{Real-Time Tracking of Switching Point $\mathbf{b_i^{SP}(t)}$ and Cycle Evolution}



\begin{figure}[t]
    \centering
    \begin{subfigure}{0.48\columnwidth}
        \centering
        \includegraphics[width=\linewidth]{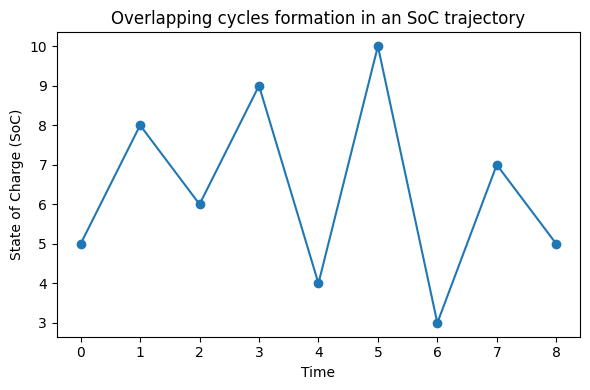}
        \caption{Overlapping cycle formation} 
        \label{fig:overlapping_cycles}
    \end{subfigure}
    \begin{subfigure}{0.48\columnwidth}
        \centering
        \includegraphics[width=\linewidth]{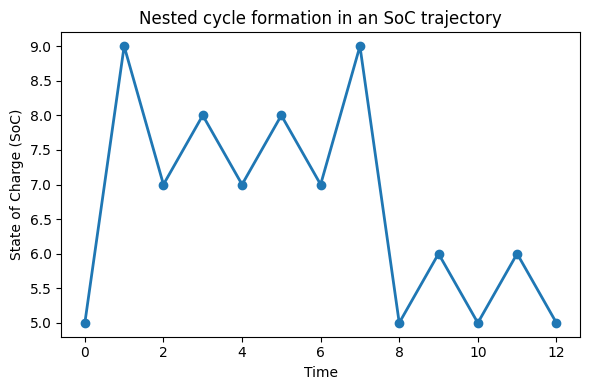}
        \caption{Nested cycle formation}
        \label{fig:nested_cycles}
    \end{subfigure}\hfill
    
    \caption{Illustration of overlapping and nested cycle formation in battery state-of-charge trajectories.}
    \label{fig:cycle_examples}
\end{figure}

The reward design in Section~3.4 relies on the most recent switching
point $b_i^{SP}(t)$ of each battery’s state-of-charge (SoC) trajectory.
While switching points are defined as local extrema of the SoC profile,
their evolution over time is more subtle than a simple forward update.
In particular, when cycles overlap or gets nested, the effective
switching point relevant for degradation assessment may move backward
along the SoC trace.

To understand this behavior, recall that charge--discharge cycles are
formed between successive extrema of the SoC trajectory. In practical
operation, cycles of different amplitudes may coexist: a smaller
cycle can be initiated and completed within a larger excursion that
has not yet closed. When such an inner cycle completes, the extrema
that define it are removed from consideration, while the larger,
enclosing cycle continues to evolve.

As a consequence, when a completed cycle is removed, the most recent
relevant switching point for the remaining incomplete cycle may
correspond to an earlier extremum along the SoC trajectory. In this
sense, the switching point $b_i^{SP}(t)$ associated with the ongoing
cycle can move backward in time as cycles are completed and discarded.
This backward movement is a natural outcome of cycle extraction and
is essential for correctly tracking the partial depth of the remaining
cycle.

For real-time operation, it is therefore necessary to maintain a
persistent list of switching points detected along the SoC trajectory
up to the current time. At each time step, newly observed SoC values
may generate new extrema, which are appended to this list. When a cycle
completes, the extrema defining that cycle are removed, potentially
exposing an earlier switching point as the active reference for the
ongoing cycle. The current switching point $b_i^{SP}(t)$ is then defined
as the most recent extremum associated with the incomplete cycle that
remains after this removal process.

Importantly, although this procedure is inspired by cycle-counting
principles, it operates incrementally and does not require access to
future SoC values. At every time step, the maintained switching-point
structure provides the correct reference $b_i^{SP}(t)$ needed to
compute the one-step MDP reward in \eqref{eq:reward_fun}-\eqref{eq:MDP_reward}. In this way, the reward
reflects the instantaneous growth of the active cycle, even as the
underlying cycle structure evolves dynamically over time.

During the performance evaluation of the trained agent, we use complete offline SoC trajectories and standard cycle-counting methods. During learning and control,
the switching-point tracking mechanism serves solely to support the
construction of a dense, real-time reward that remains aligned with
cycle-based degradation.

\section{Degradation-Aware Scheduling via Reinforcement Learning}

In this section, we present the reinforcement learning (RL) methods used to obtain degradation-aware scheduling policies. We first describe a tabular $Q$-learning approach, which is suitable for small problem instances, and then introduce a scalable function-approximation-based method using Extreme Learning Machines (ELMs) for large state–action spaces.

\subsection{Tabular Q-Learning for Small-Scale Systems}
For finite MDPs, $Q$-learning is a classical reinforcement learning algorithm.  
Let $\pi$ denote a stationary policy that maps states to feasible actions. The action-value function associated with policy $\pi$ is defined as

\begin{equation*}
    Q^\pi(s,a) = \mathbbm{E} \Big[ \sum_{t=o}^\infty \gamma^k R\big(s(t), \pi (s(t)), s(t+1) \big) \Big | s(0)=s, a(0)=a\Big]. 
\end{equation*}

The optimal action-value function $Q^\star (s,a)$
 satisfies the Bellman optimality equation (see \cite{sutton2018reinforcement}), given by

 \begin{equation*}
     Q^\star(s,a) = \mathbbm{E} \Big[R(s,a,s^\prime) + \gamma \sum_{s^\prime \in \mathcal{S}} P(s^\prime|s,a) \max_{a^\prime \in \mathcal{A}_{s^\prime}} Q^\star(s^\prime, a^\prime) \Big].
\end{equation*}
where $s^\prime$ is the successor state resulting from applying action $a$ in state $s$ and $P(s^\prime|s,a)$ is the probability of this state transition under the influence of action $a.$ This recursive relationship facilitates learning of optimal policies in finite MDPs. 




Since the transition probabilities of the regulation signal are unknown, we employ model-free Q-learning, which iteratively updates Q-values using observed transitions:
\begin{align}\label{eq:Q-learning_alg}
    Q\big(s(t), a(t)\big) \leftarrow Q(s(t), &a(t)) + \alpha(t) \Bigg\{ R\big(s(t), a(t), s(t+1)\big) + \nonumber \\
     &\gamma \max_a Q\big(s(t+1),a\big) - Q\big(s(t),a(t)\big) \Bigg\}
\end{align}
where $\alpha(t) \in (0,1)$ is the learning rate.

The conversion of $Q(s,a)$ to optimal Q-values $Q^\star (s,a),$ using the above update\footnote{with an exploratory policy that ensures every state-action pair is sufficiently explored}, is guaranteed if the learning rate $\alpha(\cdot)$ satisfies following classical Robbins–Monro conditions for stochastic approximation (see \cite{sutton2018reinforcement})
\begin{equation}\label{eq:Robbins-Monro conditions}
    \sum_t \alpha(t)=\infty, \quad \sum_t \alpha^2(t) < \infty .
\end{equation}

The converged Q functions or the Q-values obtained using \eqref{eq:Q-learning_alg} are then used to infer the RL policy as
\begin{equation*}
    \pi_{\text{Q learning}}(s) = \argmax_{a \in \mathcal{A}} Q(s,a).
\end{equation*}

\subsection{Limitations of Tabular Q-Learning}

Although conceptually simple, tabular $Q$-learning becomes impractical in our setting when, battery capacities are large,energy is discretized finely, or the number of batteries increases. In fact, the number of state–action pairs grows exponentially with $N$\footnote{recall from MDP model in Section~\ref{subsec:mdp-model} that the cardinality of state space alone grows as $|\mathcal{S}| = \left( \prod_{i=1}^{N} (B_i + 1) \right) |\mathcal{S}_r|.
$}, making storage and convergence of the $Q$-table computationally infeasible. This motivates the use of function approximation wherein the aim is to approximate the $Q$-function via some function approximation technique.

\subsection{Function Approximation for Large State–Action Spaces}
To address the curse of dimensionality, we approximate the action-value function using a linear function approximator:
\begin{equation}
\hat{Q}(s,a;w) = w^{\top}\phi(s,a),
\label{eq:linear_q}
\end{equation}
where $\phi(s,a) \in \mathbb{R}^d$ is a feature vector extracted from the state--action pair, and
$w \in \mathbb{R}^d$ is a parameter vector to be learned.

The parameters are updated using a stochastic semi-gradient method:
\begin{equation}
w \leftarrow w + \alpha_t
\Bigl(
R(s,a,s')
+ \gamma \max_{a' \in \mathcal{A}(s')}
\hat{Q}(s',a';w)
- \hat{Q}(s,a;w)
\Bigr)\phi(s,a).
\label{eq:sg_update}
\end{equation}

The quality of approximation and hence the learning performance depend critically on the choice of feature map
$\phi(\cdot,\cdot)$, which we describe next.

\subsection{Feature Extraction via Extreme Learning Machines}
\label{subsec:elm_features}

Designing effective nonlinear features for reinforcement learning is challenging, particularly in systems with strong coupling between actions and state evolution, such as battery scheduling. To capture these nonlinearities, deep neural networks can be used, but they introduce training instability and significant computational overhead. To balance expressiveness and stability, we employ Extreme Learning Machines (ELMs) as nonlinear feature extractors \cite{HUANG2006489}.

\subsubsection{ELM Architecture}
An ELM is a single-hidden-layer feedforward neural network in which the input-layer weights and biases are randomly initialized and fixed, and only the output-layer weights are learned.

Let $x \in \mathbb{R}^p$ denote the input vector to the ELM. In our setting, we use
\begin{equation}
x = [\, s,\, a, \, r, \, b^{SP}_i(t)],
\label{eq:elm_input}
\end{equation}
i.e., the concatenation of the state vector, action vector, current reward obtained, and the vector of most recent switching points across $N$ battery units, resulting in $p=3N+2$.

The hidden-layer output is given by
\begin{equation}
\phi(s,a) = \sigma(Wx + \tilde b),
\label{eq:elm_features}
\end{equation}
where $W \in \mathbb{R}^{d \times p}$ is a randomly initialized weight matrix, $\tilde b \in \mathbb{R}^d$ is a bias vector, $\sigma(\cdot)$ is an elementwise activation function (e.g., sigmoid or ReLU), and $d$ is the number of hidden units.

The hidden layer output of the ELM, $\phi(s,a)$, serves as the feature vector in the linear Q-function approximation in~\eqref{eq:linear_q}.

\subsubsection{Why ELMs Are Suitable in This Setting}
The use of ELM-based features is well suited to the present setting for several reasons. Random projections combined with nonlinear activation map state–action pairs into a high-dimensional feature space in which linear value-function approximation is effective. Because the input-layer weights are fixed, learning reduces to a linear stochastic approximation problem, avoiding the instability often associated with backpropagation-based reinforcement learning. In addition, feature computation is inexpensive and parameter updates scale linearly with the number of features, making the approach computationally efficient.

Additionally, Theorem~2.2 in~\cite{HUANG2006489} establishes that for a
single-hidden-layer network with an infinitely differentiable
activation function and randomly initialized input-layer weights and
biases, the induced ELM feature map is universal in the following
sense: for any finite dataset $\{(s_i,a_i)\}_{i=1}^M$ with targets
$Q_{\mathrm{target}}(s_i,a_i)$ and any $\varepsilon>0$, there exists a
set of output weights $w$ such that
$
\frac{1}{M}\sum_{i=1}^M
\bigl(
Q_{\mathrm{target}}(s_i,a_i)
-
w^\top \phi_{\mathrm{ELM}}(s_i,a_i)
\bigr)^2
< \varepsilon.
$
 Although derived in a supervised learning context, this result
provides theoretical justification for using randomized nonlinear
feature maps together with linear value-function approximation in the
present reinforcement learning setting.

\subsubsection{Initialization and Parameters}

The weights $W$ and biases $\tilde b$ are initialized independently and identically distributed from a uniform distribution over $[-1,1]$. The output weights $w$ are initialized to zero. The feature dimension $d$ is chosen to balance approximation accuracy and computational complexity.

After training (elaborated in Section~\ref{sec:RL_training}), the function-approximation-based policy is given by
\begin{equation}
\pi_{\mathrm{FA}}(s)
=
\arg\max_{a \in \mathcal{A}(s)}
\hat{Q}(s,a;w).
\label{eq:fa_policy}
\end{equation}

This policy selects, at each state, the feasible action satisfying \eqref{eq:ramp and feas constraints},\eqref{eq:action_constraint1},\eqref{eq:regulation_constraint} that maximizes the estimated long-term degradation-aware value.

\subsection{RL Training Algorithm}\label{sec:RL_training}

We now describe the training procedure used to learn the
function-approximation-based degradation-aware scheduling policy.
The learning algorithm builds on the semi-gradient temporal-difference
update in \eqref{eq:sg_update}, combined with experience replay and minibatch
updates to improve stability and sample efficiency.

\paragraph{Experience Collection and Replay Buffer.}
During training, the agent interacts with the environment by observing
state transitions of the form $(s(t), a(t), R(t), s(t+1))$. Each such
transition is stored in a replay buffer. The replay buffer is used to
randomly sample past experiences when updating the value-function
parameters, thereby reducing temporal correlations between successive
updates and improving the stability of learning.

\paragraph{Mini-batch Semi-Gradient Updates.}
Rather than updating the parameter vector $w$ using a single transition,
we employ mini-batch updates. At each update step, a batch of
$N_{\text{batch}}$ transitions is sampled uniformly at random from the
replay buffer. The parameter vector is then updated by averaging the
temporal-difference (TD) errors across the batch:
\begin{align}
w \leftarrow w + 
\frac{\alpha_t}{N_{\text{batch}}}
\sum_{i=1}^{N_{\text{batch}}}
\Big(
R(s_i,a_i,s'_i)
+ &\gamma \max_{a' \in \mathcal{A}(s'_i)} \hat{Q}(s'_i,a';w)
\nonumber \\
&- \hat{Q}(s_i,a_i;w)
\Big)
\phi(s_i,a_i).
\end{align}
Batch updates reduce the variance of parameter updates and mitigate the
effect of noisy individual transitions.

\paragraph{Exploration Strategy.} \label{sec:exploration_policy}
To ensure sufficient exploration of the state action space, actions
are selected using an $\epsilon$-greedy policy during training. With
probability $\epsilon_t$, the agent selects a feasible action uniformly
at random from $\mathcal{A}_{s(t}$, and with probability $1-\epsilon_t$,
it selects the action that maximizes the estimated action-value
function. The exploration parameter $\epsilon_t$ decays over time
according to a predefined schedule to gradually shift from exploration
to exploitation.

\paragraph{Learning Rate and Practical Considerations.}
The learning rate $\alpha_t$ is chosen to decrease over time to promote
stable learning. While classical Robbins–Monro conditions in \eqref{eq:Robbins-Monro conditions} provide
theoretical guidance for stochastic approximation, we note that the
use of function approximation and experience replay prohibits strict
convergence guarantees \cite{puterman2014markov,sutton2018reinforcement}. Nevertheless, empirically decaying step sizes
are observed to yield stable learning behavior in our setting (see Section~\ref{sec:exp_evaluation}).

\paragraph{Input Normalization and Activation Function.}
The inputs to the ELM feature extractor—comprising regulation signal,
battery SoC levels and actions—are normalized to comparable
scales to improve numerical conditioning and learning stability. We
find that such normalization of input is quite common in neural network-based learning, which significantly improves performance \cite{bishop2006pattern}. Among
various activation functions tested, the Sigmoid Linear Unit (SiLU)
consistently yields better empirical performance than ReLU in our
experiments, likely due to its smoother nonlinearity. Wider literature also finds the Sigmoid Linear Unit to show better performance while approximating continuous functions \cite{srinivasa2025solvinginfinitehorizonoptimalcontrol,dung_6_2023_1803}.

\paragraph{Training and Policy Extraction.}
Training is performed over multiple episodes, each corresponding to a
long realization of the regulation signal. After training converges,
the learned parameter vector $w$ is fixed, and the degradation-aware
policy is obtained by greedily selecting, at each state, the feasible
action that maximizes the approximated action-value function:
\[
\pi_{\text{FA}}(s) = \arg\max_{a \in \mathcal{A}(s)} \hat{Q}(s,a;w).
\]
The specific hyperparameter values and training configurations used in
the experiments are reported in Section~5.

\begin{table*}
\centering
\caption{Performance comparison under the toy Markov regulation model 
($\mathcal{S}_r=\{-4,-1,1,5\}$, $T=10^5$, $c_1=d_1=2$, $c_2=d_2=3$). Accumulated rewards and per-battery
cycling degradation $(D_1,D_2)$ are reported for each policy.}
\begin{tabular}{|c|c|cc|cc|cc|}
\hline
\multirow{2}{*}{\shortstack{\\ \\ $(B_1,B_2)$}} &
\multirow{2}{*}{$|\mathcal{S}\times \mathbf{A}|$} &
\multicolumn{6}{c|}{Policy Performance} \\
\cline{3-8}
 & &
\multicolumn{2}{c|}{Naive Policy} &
\multicolumn{2}{c|}{Greedy Policy} &
\multicolumn{2}{c|}{RL--ELM Policy} \\
\cline{3-8}
 & &
Reward & Degradation $(D_1,D_2)$ &
Reward & Degradation $(D_1,D_2)$ &
Reward & Degradation $(D_1,D_2)$ \\
\hline
(2,3) & 420 & -14.17 & (0.17, 0.17) & -3.89  & (0.05, 0.05) & -1.66 & (0.02, 0.0198) \\

(3,5) & 840 & -41.64 & (0.16, 0.15) & -28.41 & (0.18, 0.10) & -24.57 & (0.19, 0.1567) \\


(10,10) & 4235 & -50.46 & (0.17, 0.13) & -47.27 & (0.24, 0.23) & -44.88 & (0.10, 0.0575) \\

(15,10) & 6160 & -47.27 & (0.15, 0.13) & -44.34 & (0.11, 0.29) & -38.32 & (0.09, 0.0536) \\

(20,20) & 15435 & -29.77 & (0.15, 0.11) & -27.28 & (0.19, 0.19) & -26.01 & (0.07, 0.0406) \\

(25,25) & 23660 & -23.97 & (0.15, 0.11) & -21.95 & (0.17, 0.18) & -20.74 & (0.07, 0.0398) \\

(30,30) & 33635 & -20.31 & (0.15, 0.11) & -18.81 & (0.18, 0.17) & -17.09 & (0.07, 0.0242) \\

(40,40) & 58835 & -15.16 & (0.16, 0.11) & -13.94 & (0.17, 0.17) & -12.58 & (0.06, 0.0296) \\

(50,50) & 91035 & -12.23 & (0.15, 0.11) & -11.34 & (0.19, 0.17) & -9.90  & (0.07, 0.0348) \\
\hline
\end{tabular}
\label{table:sim_res_toy_model}
\end{table*}

\section{Experimental Evaluation}\label{sec:exp_evaluation}

In this section, we evaluate the proposed degradation-aware reinforcement learning (RL) framework using both synthetic and real-world regulation signals. The evaluation focuses on two questions: Does the learned policy reduce cycle-based battery degradation compared to reward-agnostic baselines, and can the proposed function-approximation-based RL method scale to realistic problem sizes? We answer these questions by comparing accumulated rewards and post hoc cycle statistics obtained via rainflow counting.

\subsection{Experimental Setup}

\subsubsection{Time Horizon and Granularity}
All experiments are conducted over long horizons to ensure sufficient cycle formation. Unless stated otherwise, the time horizon is $T = 10^{5}$ time steps. Energy is measured with fine granularity so that all state and action variables are integer-valued.

\subsubsection{Battery Configuration}
We consider systems with $N = 2$ heterogeneous batteries. Each experiment specifies: battery capacities $(B_1, B_2)$, charge and discharge limits $(c_i, d_i)$, and a fixed control interval duration $T$ steps. While the experiments focus on two batteries for clarity, the learning framework directly generalizes to larger $N$.

\subsubsection{Agents Compared}

To assess the benefits of degradation-aware learning/training, we compare the proposed reinforcement learning agents against two baseline agents that do not explicitly consider cycling degradation. 

\paragraph{Reward-Agnostic Naive Agent}
A simple baseline that allocates the regulation signal in proportion to battery capacities:
\begin{equation}
a_i(t)
=
\left\lfloor
\frac{B_i}{\sum_{j=1}^{N} B_j}\, r(t)
\right\rfloor,
\label{eq:capacity_proportional}
\end{equation}
with any rounding mismatch allocated to a feasible battery so as to satisfy
\eqref{eq:ramp and feas constraints} and \eqref{eq:regulation_constraint}.
Specifically, the remaining energy resulting from the floor operation is allocated, one unit at a time, to feasible batteries selected uniformly at random.
At each allocation step, the feasible set consists of batteries that satisfy the ramp and state-of-charge constraints
\eqref{eq:ramp and feas constraints}, \eqref{eq:regulation_constraint} and have sufficient headroom (for charging) or available energy (for discharging).

\paragraph{Greedy Agent}
Another myopic policy that chooses the action based on one one-step MDP reward as
\begin{equation}
   a(t)=\max_{a(t)\in \mathcal{A}_{s(t)}} R(s(t),a(t),s(t+1)) 
\end{equation}


These two baselines serve as reference points for evaluating the benefits of learning degradation-aware policies. We compare these baselines with two RL agents--tabular Q-learning (only for small state spaces) and a function approximation-based RL agent (ELM-RL) that uses ELM-based feature extraction and linear $Q$-function approximation.
\paragraph{RL training details}
The RL agents use an $\epsilon$-greedy exploration strategy during training, with a decaying exploration rate as 
$
    \varepsilon_t = \frac{\varepsilon}{1+0.005t},
$
with $\varepsilon=0.6$. 
As mentioned in Section~\ref{sec:RL_training}, the learning rate  $\alpha_t$ is varied as $\alpha_{t} = \frac{\alpha} {N_t(s,a)}$, where $\alpha = 10^{-4}$ and $N_t(s,a)$ is the number of times the currently encountered state-action pair $(s,a)$ has been visited till time step $t.$ A minibatch of size $N_{batch}=128$ is used to update the weights $w$ every 8 samples or interaction with the environment during RL training. This minibatch is sampled from the reply buffer whose size is kept as 2000 for the toy setting and 160,000 for the simulation with PJM trace. 

\subsubsection{Performance Metrics}

We report the accumulated reward
$
R^{\pi}
=
\sum_{t=0}^{T-1}
R\bigl(s(t), a(t), s(t+1)\bigr),
$
 along the sample path of the evolved state trajectory $\{s(t)\}_{t=0}^{T-1}$ over which we want to evaluate the performance of the RL agent and other baseline agents. We also report cycle statistics obtained from rainflow counting, including: number of cycles, depth-of-discharge (DoD) histograms, total degradation computed using the stress function in ~\eqref{eq:tot_degradation_in_bat_i}.\footnote{We use the stress function $(k_{\delta_1}\delta^{k_{\delta_2}} + k_{\delta_3})^{-1}$  with $k_{\delta_1}=1.4\times 10^5$, $k_{\delta_2}=-5.01\times 10^{-1}$, and $k_{\delta_3}=-1.23 \times 10^5$. These constants can be found in \cite{bat_degradation1}.}  These latter metrics are computed offline using the full state-of-charge trajectories and represent the true performance objective.

\begin{figure*}[t]
    \centering
    \includegraphics[width=0.75\linewidth,trim=10pt 8pt 10pt 6pt,
  clip]{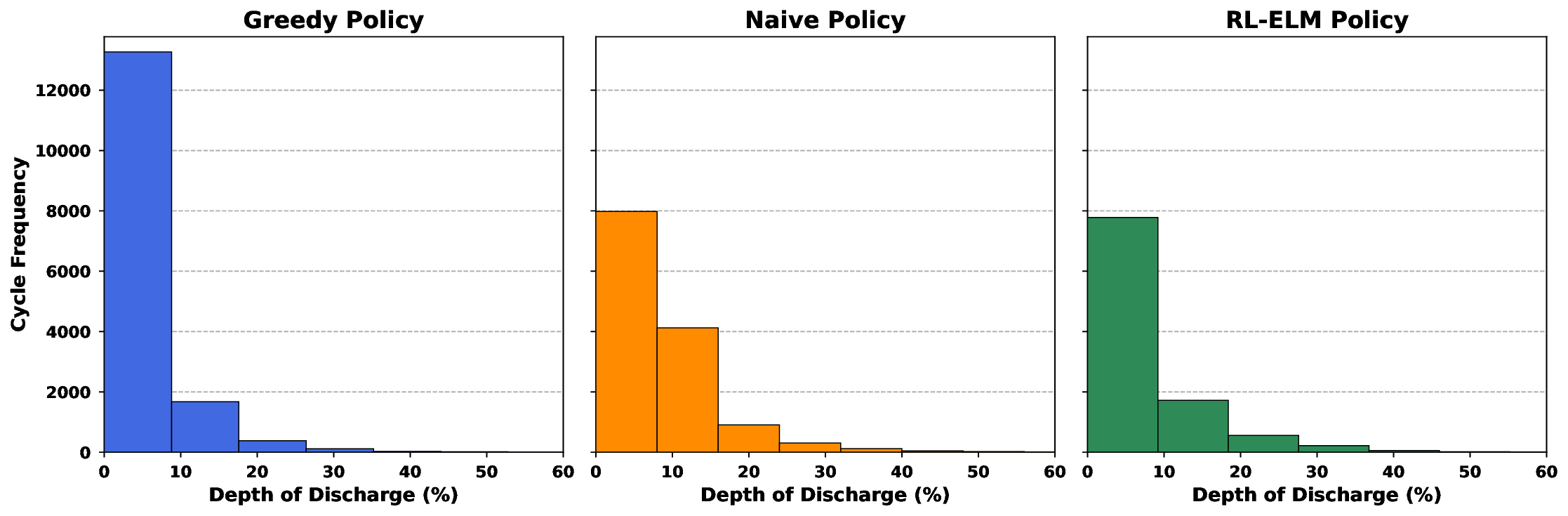}
     \caption{Distribution of cycle occurrence across DoD levels under different control policies in battery of size $B=25$.}
    \label{fig:ToyModel_Bat2_cycles_in_a_2_BatterySetup}
\end{figure*}

\begin{figure*}[t]
    \centering
    \includegraphics[width=0.75\linewidth, trim=10pt 8pt 10pt 6pt,
  clip]{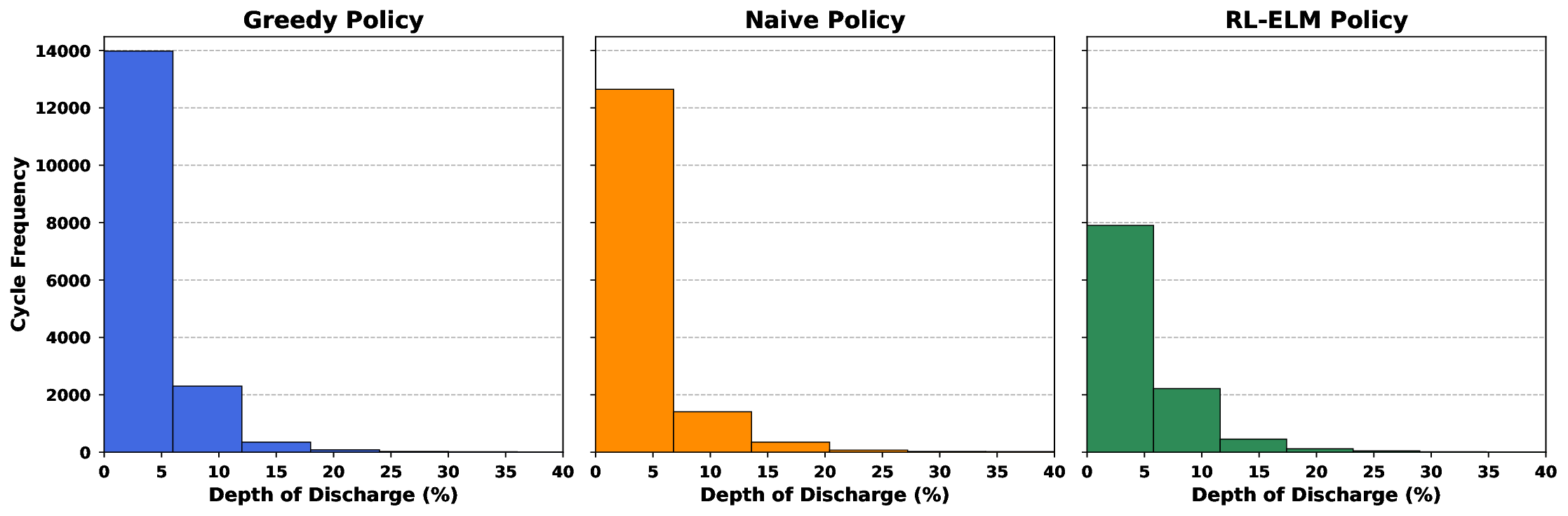}
    \caption{Distribution of cycle occurrence across DoD levels under different control policies in battery of size $B=50$.}
    \label{fig:ToyModel_Bat2_cycles_in_a_2_BatterySetup_}
\end{figure*} 

\subsection{Synthetic Regulation Signal (Toy Markov Model)}

We first evaluate the agents using a small synthetic discrete-time Markov chain (DTMC) to validate learning behavior under controlled conditions.

\subsubsection{Signal Model}

The regulation signal evolves over the finite state space
$
S_r = \{-4, -1, 1, 5\},
$
with a fixed transition probability matrix. The transition matrix is used only to generate the signal trajectory and is unknown to all agents.

\subsubsection{Battery Parameters}

We consider two batteries with identical ramp limits,
\begin{equation}
c_1 = d_1 = 2, \, c_2 =d_2 = 3,
\end{equation}
and vary the capacities $(B_1, B_2)$ across experiments.

\subsubsection{Results}
For various battery size configurations $(B_1, B_2),$ each defining an MDP instance, the accumulated reward obtained with RL agents and with other baseline naive agents is reported in \Cref{table:sim_res_toy_model}.
Across all battery size configurations (see Table~\ref{table:sim_res_toy_model}), the reinforcement learning agents outperform the naive baseline in terms of accumulated reward. The function-approximation-based RL agent consistently achieves the highest reward, indicating lower degradation. To interpret these results physically, we examine rainflow-counted cycle statistics. As reported in Table~\ref{table:sim_res_toy_model}, the actual cycling degradation computed offline (after the simulation using the full SoC trajectory) using the rainflow counting algorithm is less in both battery units compared to the operation carried out by the baseline naive and greedy policies. The learned RL policy reduce the number of deep cycles, redistribute regulation effort across batteries, and induce more shallow cycling behavior. Histograms of cycle depth (see Fig.~\ref{fig:ToyModel_Bat2_cycles_in_a_2_BatterySetup} and \ref{fig:ToyModel_Bat2_cycles_in_a_2_BatterySetup_}) show a clear reduction in both the frequency and magnitude of high depth-of-discharge (DoD) cycles under RL control. These trends directly explain the higher accumulated reward and lower total degradation. 

\subsection{Evaluation with PJM RegD Frequency Regulation Signal}\label{sec:pjm_regd}
To validate performance under a realistic regulation signal, we evaluate the proposed framework using historical PJM RegD data. The data was collected
during the University of Delaware vehicle-to-grid (V2G) field
demonstration in Spring 2013 (see here \cite{UD_V2G_FastReg2013}). The dataset consists of a high-resolution
RegD signal sampled at 10-second intervals and reflects the fast,
zero-mean regulation service request deployed by PJM.

\subsubsection{Experimental Details}
\label{subsec:pjm_regd_experiment}

The RegD signal is normalized with respect to a regulation capacity of
$A_{\mathrm{Reg}} = 10$~kWh, such that a normalized value of $+1$
corresponds to a request for $+10$~kWh of charging over a control
interval, while a value of $-1$ corresponds to $-10$~kWh of
discharging. We sample the RegD signal at a resolution of 0.1, which translates to 21 possible regulation values over the trace. This stochastically varying RegD signal, spanning $109{,}846$ samples, serves as the training and evaluation signal for the RL agent. This trajectory is treated as a single episode,
allowing the learning agent to experience realistic long-term cycling
behavior without artificial episode resets.



\subsubsection{Results}
\label{subsec:pjm_regd_results}

Tables~\ref{table:sim_res_pjm_50} report performance under the PJM RegD signal for different
battery configurations across different policies.
Across all MDP instances, the RL--ELM agent consistently
outperforms the naive and greedy baselines in terms of accumulated
reward and rainflow-counted cycling degradation. The learned policies
exhibit a marked reduction in deep cycles and a more balanced
distribution of regulation effort across the battery units. Note that  even with as few as 10 hidden units, the RL--ELM agent maintains a substantial advantage over baseline policies.

\begin{table*}[t]
\centering
\caption{Performance comparison under the PJM Regulation Data
($c_1=d_1=5$, $c_2=d_2=10$, ELM feature dimension $d=50.$). Accumulated rewards and per-battery cycling degradation $(D_1,D_2)$ are reported for each policy.}
\begin{tabular}{|c|c|cc|cc|cc|}
\hline
\multirow{2}{*}{\shortstack{\\ \\ $(B_1,B_2)$}} &
\multirow{2}{*}{$|\mathcal{S}\times \mathbf{A}|$} &
\multicolumn{6}{c|}{Policy Performance} \\
\cline{3-8}
 & &
\multicolumn{2}{c|}{Naive Policy} &
\multicolumn{2}{c|}{Greedy Policy} &
\multicolumn{2}{c|}{RL--ELM Policy} \\
\cline{3-8}
 & &
Reward & Degradation $(D_1,D_2)$ &
Reward & Degradation $(D_1,D_2)$ &
Reward & Degradation $(D_1,D_2)$ \\
\hline
(5, 10) & 15,246 & -5.89 & (0.003732, 0.006304) & -4.64 & (0.003516, 0.003785) & -4.64 & (0.003516, 0.003785) \\

(10, 10) & 27,951 & -7.01 & (0.006284, 0.006208) & -5.81 & (0.004136, 0.00526) & -5.79 & (0.004295, 0.005343) \\

(15, 10) & 40,656 & -7.44 & (0.007506, 0.006245) & -6.46 & (0.004689, 0.005918) & -6.46 & (0.004689, 0.005918) \\

(15, 20) & 77,616 & -7.96 & (0.007325, 0.008113) & -7.18 & (0.007067, 0.005162) & -7.18 & (0.007067, 0.005162) \\

(20, 20) & 101,871 & -8.12 & (0.008053, 0.008087) & -7.37 & (0.00603, 0.006304) & -7.37 & (0.005754, 0.006407) \\

(20, 30) & 150,381 & -8.20 & (0.008029, 0.008635) & -7.60 & (0.00742, 0.005628) & -7.60 & (0.00742, 0.005628) \\

(30, 30) & 221,991 & -8.26 & (0.008608, 0.008592) & -7.73 & (0.0064, 0.006215) & -7.73 & (0.006086, 0.006176) \\

(50, 50) & 600,831 & -8.04 & (0.008742, 0.008436) & -7.69 & (0.005412, 0.005349) & -7.67 & (0.004521, 0.004756) \\

(75, 75) & 1,334,256 & -7.56 & (0.008161, 0.008317) & -7.31 & (0.004549, 0.004449) & -7.28 & (0.003969, 0.00408) \\

(75, 100) & 1,773,156 & -7.28 & (0.00762, 0.008007) & -7.06 & (0.006712, 0.003731) & -7.04 & (0.006107, 0.003499) \\

(100, 100) & 2,356,431 & -7.03 & (0.007622, 0.007455) & -6.84 & (0.004554, 0.004119) & -6.82 & (0.003444, 0.003306) \\

\hline
\end{tabular}

\label{table:sim_res_pjm_50}
\end{table*}

\section{Conclusion, Limitations, and Future Work}

This paper studied the problem of degradation-aware frequency regulation
using a fleet of heterogeneous battery storage units. By formulating the
control problem as a Markov decision process with a carefully designed
proxy reward, we demonstrated that reinforcement learning can
systematically shape state-of-charge trajectories to reduce long-term
cycling degradation under stochastic regulation signals. Extensive
numerical experiments using both synthetic and real-world regulation
data confirm that the proposed approach consistently suppresses deep
cycles and lowers degradation compared to baseline dispatch policies.

Several limitations of the present work point to promising directions
for future research. While frequency regulation tracking
requirements are enforced through hard feasibility constraints on the
action space, explicit penalties for regulation error are not directly
incorporated into the MDP reward. A natural extension is to integrate
tracking performance and degradation costs within a unified reward
structure, enabling a more explicit exploration of the trade-off
between regulation accuracy and battery lifetime.

The present model focuses exclusively on cycle-induced
degradation and does not explicitly account for other aging mechanisms such as calendar aging, temperature effects, or rate-dependent losses. Incorporating degradation models that capture these additional
factors, while maintaining tractability for online learning, remains an important open challenge.

\bibliographystyle{ACM-Reference-Format}
\bibliography{refs}

@String{Springer = "Springer-Verlag" }

@book{sutton2018reinforcement,
  title={Reinforcement learning: An introduction},
  author={Sutton, Richard S and Barto, Andrew G},
  year={2018},
  publisher={MIT press}
}

@article{bat_degradation1,
	author={Xu, Bolun and Oudalov, Alexandre and Ulbig, Andreas and Andersson, Göran and Kirschen, Daniel S.},
	journal={IEEE Transactions on Smart Grid}, 
	title={Modeling of Lithium-Ion Battery Degradation for Cell Life Assessment}, 
	year={2018},
	volume={9},
	number={2},
	pages={1131-1140},
	doi={10.1109/TSG.2016.2578950}
}

@INPROCEEDINGS{bat_deg_Millner,
  author={Millner, Alan},
  booktitle={2010 IEEE Conference on Innovative Technologies for an Efficient and Reliable Electricity Supply}, 
  title={Modeling Lithium Ion battery degradation in electric vehicles}, 
  year={2010},
  volume={},
  number={},
  pages={349-356},
  doi={10.1109/CITRES.2010.5619782}}

@ARTICLE{bat_deg_via_RL,
  author={Cao, Jun and Harrold, Dan and Fan, Zhong and Morstyn, Thomas and Healey, David and Li, Kang},
  journal={IEEE Transactions on Smart Grid}, 
  title={Deep Reinforcement Learning-Based Energy Storage Arbitrage With Accurate Lithium-Ion Battery Degradation Model}, 
  year={2020},
  volume={11},
  number={5},
  pages={4513-4521},
  doi={10.1109/TSG.2020.2986333}}

@article{downing1982simple,
  title={Simple rainflow counting algorithms},
  author={Downing, Stephen D and Socie, DF},
  journal={International journal of fatigue},
  volume={4},
  number={1},
  pages={31--40},
  year={1982},
  publisher={Elsevier}
}

@inproceedings{shi2018convex,
  title={A convex cycle-based degradation model for battery energy storage planning and operation},
  author={Shi, Yuanyuan and Xu, Bolun and Tan, Yushi and Zhang, Baosen},
  booktitle={2018 Annual American Control Conference (ACC)},
  pages={4590--4596},
  year={2018},
  organization={IEEE}
}

@article{HUANG2006489,
title = {Extreme learning machine: Theory and applications},
journal = {Neurocomputing},
volume = {70},
number = {1},
pages = {489-501},
year = {2006},
note = {Neural Networks},
issn = {0925-2312},
doi = {https://doi.org/10.1016/j.neucom.2005.12.126},
author = {Guang-Bin Huang and Qin-Yu Zhu and Chee-Kheong Siew}
}

@ARTICLE{KWONZHU9789478,
  author={Kwon, Kyung-Bin and Zhu, Hao},
  journal={IEEE Transactions on Smart Grid}, 
  title={Reinforcement Learning-Based Optimal Battery Control Under Cycle-Based Degradation Cost}, 
  year={2022},
  volume={13},
  number={6},
  pages={4909-4917},
  doi={10.1109/TSG.2022.3180674}}

@inproceedings{srinivasa2025solvinginfinitehorizonoptimalcontrol,
  title={Solving Infinite-Horizon Optimal Control Problems using the Extreme Theory of Functional Connections},
  author={Srinivasa, Tanay Raghunandan and Kumar, Suraj},
  booktitle={Proceedings of the 11th Indian Control Conference (ICC)},
  year={2025},
  note={Presented at ICC-11; also available as arXiv:2510.27187},
  url={https://arxiv.org/abs/2510.27187}
}

@inproceedings{dung_6_2023_1803,
author = {Duong V. Dung and Nguyen D. Song and Pramudita S. Palar and Lavi R. Zuhal},
title = {On The Choice of Activation Functions in Physics-Informed Neural Network for Solving Incompressible Fluid Flows},
booktitle = {AIAA SCITECH 2023 Forum},
published = {},
chapter = {},
pages = {},
doi = {10.2514/6.2023-1803},
year = {2023}
}

@misc{UD_V2G_FastReg2013,
  author        = {Kempton, Willett and {University of Delaware V2G Research Group}},
  title         = {Fast Regulation Signal Data (PJM REG-D)},
  year         = {2013},
  howpublished = {University of Delaware V2G Resources},
  url          = {https://www1.udel.edu/V2G/resources/Fast-Regulation-Data.csv},
  note         = { Accompanied by Fast-Regulation-Explanation.txt}
}

@book{bishop2006pattern,
  author    = {Bishop, Christopher M.},
  title     = {Pattern Recognition and Machine Learning},
  publisher = {Springer},
  year      = {2006},
  series    = {Information Science and Statistics},
  address   = {New York},
  isbn      = {978-0-387-31073-2}
}

@book{puterman2014markov,
	title={Markov decision processes: discrete stochastic dynamic programming},
	author={Puterman, Martin L},
	year={2014},
	publisher={John Wiley \& Sons}
}

@article{vazquez2019reinforcement,
  title={Reinforcement learning for demand response: A review of algorithms and modeling techniques},
  author={V{\'a}zquez-Canteli, Jos{\'e} R and Nagy, Zolt{\'a}n},
  journal={Applied energy},
  volume={235},
  pages={1072--1089},
  year={2019},
  publisher={Elsevier}
}

@article{mnih2015human,
  title={Human-level control through deep reinforcement learning},
  author={Mnih, Volodymyr and Kavukcuoglu, Koray and Silver, David and Rusu, Andrei A and Veness, Joel and Bellemare, Marc G and Graves, Alex and Riedmiller, Martin and Fidjeland, Andreas K and Ostrovski, Georg and others},
  journal={nature},
  volume={518},
  number={7540},
  pages={529--533},
  year={2015},
  publisher={Nature Publishing Group}
}

@article{luo2015overview,
  title={Overview of current development in electrical energy storage technologies and the application potential in power system operation},
  author={Luo, Xing and Wang, Jihong and Dooner, Mark and Clarke, Jonathan},
  journal={Applied energy},
  volume={137},
  pages={511--536},
  year={2015},
  publisher={Elsevier}
}

@article{wankmuller2017impact,
  title={Impact of battery degradation on energy arbitrage revenue of grid-level energy storage},
  author={Wankm{\"u}ller, Florian and Thimmapuram, Prakash R and Gallagher, Kevin G and Botterud, Audun},
  journal={Journal of Energy Storage},
  volume={10},
  pages={56--66},
  year={2017},
  publisher={Elsevier}
}

@article{wan2018model,
  title={Model-free real-time EV charging scheduling based on deep reinforcement learning},
  author={Wan, Zhiqiang and Li, Hepeng and He, Haibo and Prokhorov, Danil},
  journal={IEEE Transactions on Smart Grid},
  volume={10},
  number={5},
  pages={5246--5257},
  year={2018},
  publisher={IEEE}
}

@inproceedings{wang2018energy,
  title={Energy storage arbitrage in real-time markets via reinforcement learning},
  author={Wang, Hao and Zhang, Baosen},
  booktitle={2018 IEEE Power \& Energy Society General Meeting (PESGM)},
  pages={1--5},
  year={2018},
  organization={IEEE}
}

@article{hao2023adaptive,
  title={Adaptive model-based reinforcement learning for fast-charging optimization of lithium-ion batteries},
  author={Hao, Yuhan and Lu, Qiugang and Wang, Xizhe and Jiang, Benben},
  journal={IEEE Transactions on Industrial Informatics},
  volume={20},
  number={1},
  pages={127--137},
  year={2023},
  publisher={IEEE}
}

@article{zhang2024load,
  title={Load forecasting-based learning system for energy management with battery degradation estimation: A deep reinforcement learning approach},
  author={Zhang, Hongtao and Zhang, Guanglin and Zhao, Mingbo and Liu, Yuping},
  journal={IEEE Transactions on Consumer Electronics},
  volume={70},
  number={1},
  pages={2342--2352},
  year={2024},
  publisher={IEEE}
}

@article{zhang2020identifying,
  title={Identifying degradation patterns of lithium ion batteries from impedance spectroscopy using machine learning},
  author={Zhang, Yunwei and Tang, Qiaochu and Zhang, Yao and Wang, Jiabin and Stimming, Ulrich and Lee, Alpha A},
  journal={Nature communications},
  volume={11},
  number={1},
  pages={1706},
  year={2020},
  publisher={Nature Publishing Group UK London}
}

@article{liu2013decentralized,
  title={Decentralized vehicle-to-grid control for primary frequency regulation considering charging demands},
  author={Liu, Hui and Hu, Zechun and Song, Yonghua and Lin, Jin},
  journal={IEEE Transactions on Power Systems},
  volume={28},
  number={3},
  pages={3480--3489},
  year={2013},
  publisher={IEEE}
}

@article{janjic2017commercial,
  title={Commercial electric vehicle fleet scheduling for secondary frequency control},
  author={Janjic, Aleksandar and Velimirovic, Lazar and Stankovic, Miomir and Petrusic, Andrija},
  journal={Electric Power Systems Research},
  volume={147},
  pages={31--41},
  year={2017},
  publisher={Elsevier}
}

\end{document}